# Topological transitions and surface umklapp scattering in Slack Metasurfaces

**Kobi Cohen,[1] Shai Tsesses,[1] Shimon Dolev,[1] Yael Blechman,[1] Guy Ankonina,[2] and Guy Bartal[1,*]**

[1] *The Andrew and Erna Viterbi faculty of Electrical and Computer Engineering, Technion- Israel Institute of Technology, Haifa, Israel*
[2] *The photovoltaic lab, The Russell Berrie Nanotechnology Institute, Technion- Israel Institute of Technology, Haifa, Israel*
*\*guy@ee.technion.ac.il*

**Abstract:** Metamaterials and metasurfaces are at the pinnacle of wave propagation engineering, yet their design has thus far been mainly focused on deep-subwavelength periodicities, practically forming an effective medium. Such an approach overlooks important structural degrees-of-freedom, e.g. the interplay between the corrugation periodicity and depth and how it affects the beam transport. Here, we present Slack Metasurfaces – weakly modulated metal-dielectric interfaces unlocking all structural degrees-of-freedom that affect the wave propagation. We experimentally demonstrate control over the anisotropy of surface waves in such metasurfaces, leading to yet, unexplored, dual stage topological transitions. We further utilize these metasurfaces to show unique backward focusing of surface waves driven by an umklapp process – momentum relaxation empowered by the periodic nature of the structure. Our findings can be applied to any type of guided waves, introducing a simple and diverse method for controlling wave propagation in artificial media.

## 1. Introduction

Metasurfaces (MSs) – the interfaces between two materials structured in the subwavelength scale [1] – have been widely investigated in recent years for their ability to manipulate light for various applications. These include beam deflection [2], planar lensing [3], holography [4] or dynamic data encryption [5], using special meta-atoms [6,7]. Likewise, MSs play an important role in surface wave engineering, attaining unidirectional [8] and spin-selective coupling [9], as well as surface wave-guidance at low frequencies [10].

Interestingly, the surface wave properties can be strongly influenced even by simple periodic structuring of the interface, exhibiting tunable propagation or guidance characteristics. In this respect, one-dimensional depth modulation was shown to produce hyperbolic metasurfaces (HMSs), wherein the modulation amplitude tuned the spatial dispersion anisotropy from elliptic to hyperbolic [11–14], giving rise to a topological transition predicted by the effective medium theory (EMT) [15]. Such in-plane hyperbolic dispersion was later achieved even in natural, unmodulated materials: from bulk crystals as calcite [16] to van der Waals materials as α-MoO3 [17] ,enabling enhanced control over the anisotropy using a rotation of the atomic layers [18].

On the other hand, surface corrugations at the order of the wavelength have been studied for decades, offering analytical prediction for the dispersion of surface waves [19–21]. Nevertheless, to our knowledge, such corrugations have never been considered as an additional degree of freedom for engineering the transport properties of electromagnetic surface waves, beyond the EMT, nor have such surface modes been directly mapped. In this regime, surface wave properties are dominated by periodicity-induced effects, such as umklapp scattering – momentum relaxation in integer values of the modulation wavenumber – opening a path towards exotic wave dispersion and diffraction.

Here, we introduce the slack metasurface - a periodically modulated interface that goes beyond effective medium, exhibiting new degrees of control over surface wave transport. Our approach exhibits new intriguing transport regimes in a planar structure, inducing topological transitions that are unattainable in continuous or effective media. We further utilize phase-resolved near-field microscopy to demonstrate a novel "reversible" topological transition together with the unique phenomenon of backward focusing – mediated by umklapp scattering. Our findings offer a comprehensive approach for engineering surface and guided waves in artificial periodic systems and can be implemented to produce novel kinds of photonic circuitry and electromagnetic modes with intricate and unique dispersion features.

## 2. The slack metasurface

Metal-dielectric interfaces are known to accommodate surface plasmon polaritons (SPPs) – Transversely Magnetic (TM) surface waves. In a smooth metal-air surface, away from the surface plasmon resonance (illustrated in Figure 1a1), the electric field is almost normal to the surface, leading to isotropic SPPs in 2D space whose surface propagation is portrayed by a circular equi-frequency contour (EFC) in the spatial frequency domain (fig. 1a2). On the other hand, infinitely deep modulation of this interface results in a metal-dielectric multilayer (MDML), whose electric field is normal to the interfaces (along the Y-axis, fig. 1b1). Such MDMLs have been extensively studied in the framework of Hyperbolic Metamaterials named after their resultant hyperbolic EFC [3,22](fig 1b2).

Weaker one-dimensional modulation of a single interface (Figure 1c1) results in a hybrid plasmonic mode whose EFC is controlled by the interplay between the isotropic SPPs and the hyperbolic plasmons of the HMS, tuned by the modulation depth compared to the plasmon wavelength. This interplay manifests various propagation regimes, as for shallow modulation the EFC is elliptic (hereby referred to as the "gap-less EFC", fig. 1c2) while it appears as an open curve for deeper modulation (hereby referred to as the "gapped EFC", flat or semi-hyperbolic in fig. 1c3,4, respectively). Such EFCs were so far analyzed under the EMT, introducing HMSs that their response was tuned by the wavelength of the incident beam [11]. Going beyond this approximation unlocks the periodicity degree of freedom, which, together with the modulation depth, opens new possibilities to alter surface wave transport.

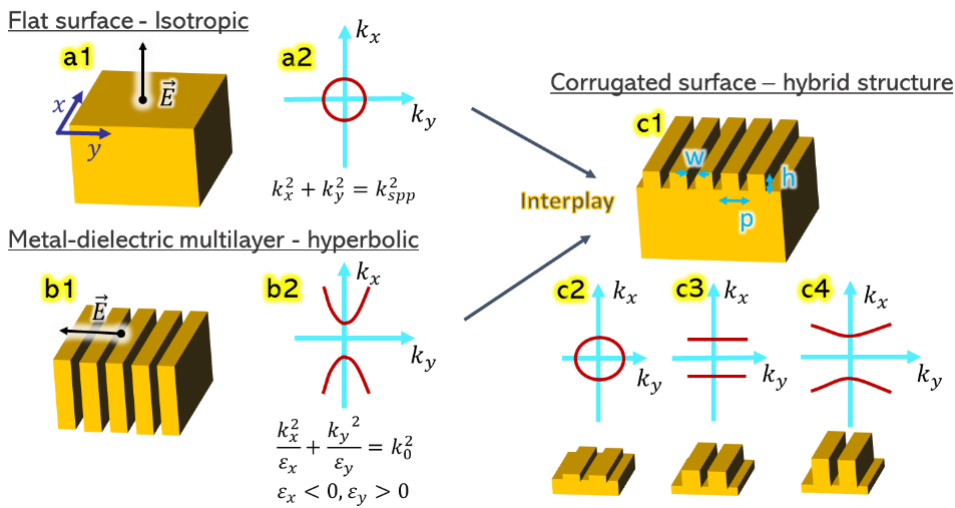

Figure 1. **Hybridizing isotropic and hyperbolic plasmons in periodically structured metasurface – schematic description**. (**a1**) A flat metal-air surface accommodates isotropic SPPs with electric field almost normal to the surface (dominant z-component), results in a circular EFC (**a2**). (**b1**) A metal-air multilayer structure supporting a bulk

mode with electric field mostly normal to the layers (dominant y-component), results in a hyperbolic EFC **(b2)**. **(c1)** A periodically modulated surface embodies the interplay between the hyperbolic and isotropic modes, while the slit depth tunes the overall response. **(c2)** Shallow grooves result in elliptic dispersion. **(c3)** Flat-band dispersion for deeper grooves. **(c4)** Hyperbolic-like dispersion at very deep grooves.

When the EMT is broken, the EFC is modified to contain an infinite array of non-circular, equally-spaced contours, divided into different Brillouin zones (BZs). The modulation period $p$ determines the separation parameter between the BZs, given by the modulation wavenumber $k_g = \frac{2\pi}{p}$, where the first BZ is located in the region $\left[-\frac{k_g}{2}, \frac{k_g}{2}\right]$. The depth $h$ typically determines the degree of anisotropy [23], such that weak modulation leads to a nearly-circular elliptic contour, while strong modulation retains the hyperbolic dispersion within each BZ. Figure 2a presents the various propagation regimes enabled by this additional degree of freedom, where, for each periodicity, the EFCs are calculated at different modulation depths.

Indeed, at small periodicities (fig. 2a1 and supplemental video 1), the EFC agrees fairly well with the effective medium theory, displaying a transformation from circular to elliptic and then hyperbolic shape [11,12,14,15,24,25], as the depth increases. This tendency is maintained at larger periodicities, (fig. 2a2) as long as the periodicity is sufficiently smaller than the wavelength, inhibiting "interaction" between neighboring zones. However, at larger periodicities (>100nm for the conditions shown herein), the effective medium approximation collapses, forming new propagation schemes: fig. 2a3 demonstrates the band stretch and gap opening resulting in a "parabolic anisotropy" when curves of two BZs merge (e.g., for p=200nm and h=50nm). A zoomed-in plot (fig. 2c) better illustrates how these parabolic curves flatten when the modulation depth is increased and become locally hyperbolic (h=130nm). As the period becomes larger than $\frac{\lambda_0}{2}$ (e.g., fig. 2a6 with p=350nm), the EFC cannot attain the gapless form and hence remains only gapped (parabolic, flat or hyperbolic).

## 3. Topological transitions in slack metasurfaces

These propagation regimes can be arranged into a "phase diagram", where the parameters of a gap-less EFC (blue), a gapped EFC with positive curvature (turquoise, denoted as "parabolic") and a gapped EFC with negative curvature (green, denoted as "hyperbolic") can be mapped and characterized, as shown in fig. 2b. The boundary between parabolic and hyperbolic EFCs corresponds to a narrow parameter space at which the dispersion curvature is either negligible (red line, denoted as "flatband"), leading to diffraction-less propagation in the direction perpendicular to the modulation; or featuring both positive and negative curvatures, depending on the propagation angle (yellow, denoted as "hybrid").

Most intriguingly, we identify new periodicity-dependent topological transitions between the gap-less and gapped EFC, emanating from two competing length-scales in the reciprocal plane: the eccentricity of the contours in each BZ and the spacing between them. The red arrow in fig. 2a marks a double transition – as the periodicity increases, the gap is closed due to insufficient eccentricity and then reopens owing to the overlap between adjacent contours. Being periodicity-dependent, such transitions could by no means be predicted by effective medium theory, which does not take the periodicity into account altogether. The double transition is illustrated in supplemental video 2 and depicted in fig. 2d; for constant grating depth, the curve is initially open (104nm), then closes (120nm) and reopens (200nm) as the periodicity grows.

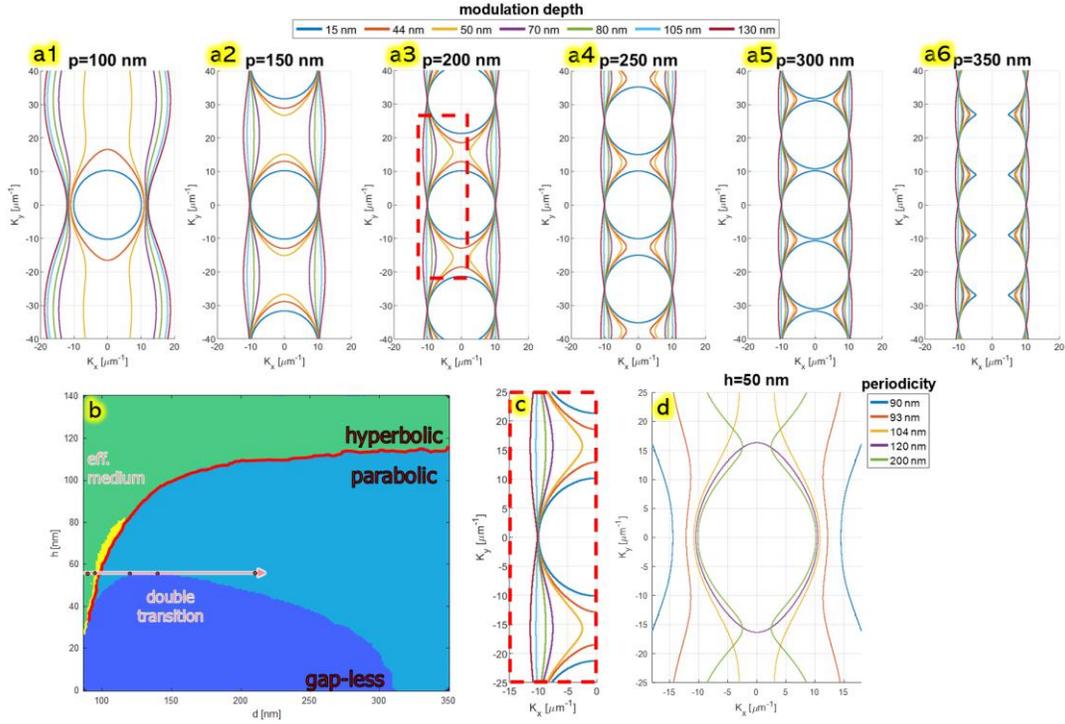

Figure 2- dispersion and topological transitions of surface waves in slack metasurface. (a1-6) EFC of slack metasurfaces at different modulation depths and periodicities, calculated using the Fourier modal method [26], supplement section 2. All plots are for constant slit width, $w = 80nm$ and free-space wavelength of $660nm$. Each plot corresponds to different periodicity and each color represents different modulation depth, outlined in the legend. Evidently, shallow grating ($h = 15nm$) results in isotropic (circular) curves for all periods. At deeper grating ($h = 44nm$) the curves become elliptic or parabolic, depending on the period. Hyperbolic curves emerge for much deeper grooves ($h = 130nm$). The transformation from a gap-less to gapped EFC occurs faster for larger periodicities (for example, the curve of $h = 44nm$ in a3,4); however, for periodicity is greater than half the excitation wavelength (e.g., p=350nm) the gap-less phase completely disappears. (b) A numerically calculated "phase diagram" illustrating the dispersion regime dependence on grating depth and periodicity, see supplement section 3. (c) A zoomed-in plot of (a3), presenting the transition from gap-less to hyperbolic according to the same legend as (a). (d) Double topological transition (gapped to gap-less to gapped), corresponding to the red arrow and dots in (b). The different colors correspond to different periodicities, noted in the inset legend. While periodicity of $p = 120nm$ results in a gapless EFC, both smaller ($104nm$) and larger ($200nm$) periodicities produce a gapped (either hyperbolic or parabolic) EFC.

We experimentally demonstrate these structurally dependent transitions and dispersion regimes by structuring slack metasurfaces using a focused-ion-beam-patterned, 200nm gold layer, deposited atop a glass substrate (fig. 3a-b). We excite the modes of the structured surface by weakly focusing a circularly polarized laser beam onto a circular slit milled through the entire gold layer. The surface-wave out-of-plane electric field distribution is collected via a phase-resolved scanning near-field microscope (s-SNOM, Neaspec), as depicted in fig. 3c. The amplitude and phase of a typical pattern, shown in fig. 3d and 3e, respectively, are used to digitally reconstruct a map of the wavevectors propagating on the surface (fig. 3f), by applying a 2D Fourier transform on the signal within the red dashed square in fig. 3d,e. This extracts the full EFC profile of a given modulated surface. The experimental results are compared to the simulation, given in fig. 3g, where the ohmic losses taken into account and the BZs marked in different colors.

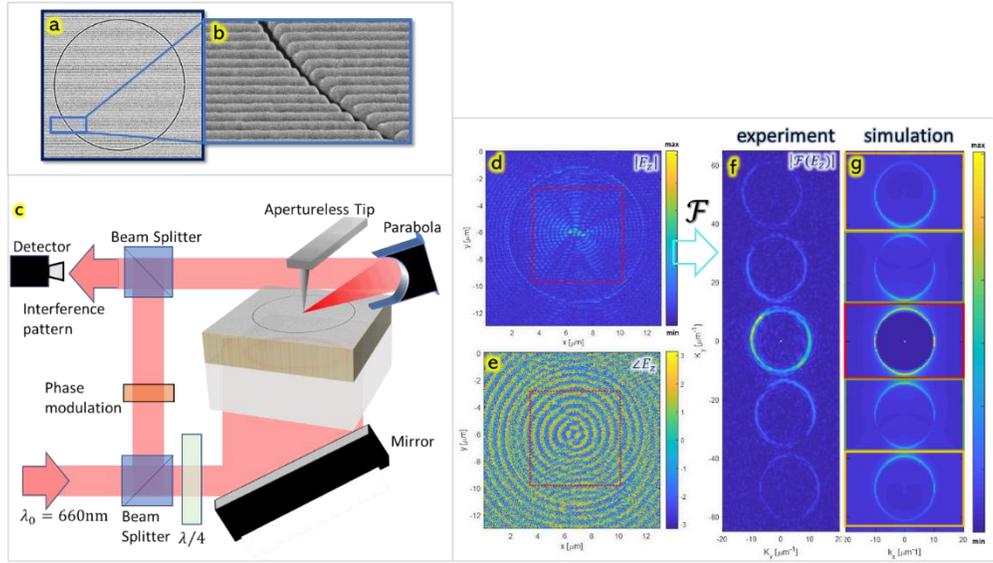

Figure 3- **Experimental setup for reconstruction of the equi-frequency contour**. (**a**) Characteristic scanning electron microscope (SEM) micrograph of the patterned gold surface used in the experiment. (**b**) Close-up of (a), emphasizing the difference between the 1D surface modulation and the circular coupling slit. (**c**) An illustration of the phase-resolved near-field microscope used for the measurement. 660 nm laser light is converted to circular polarization via a quarter-waveplate, and then impinges the sample from the glass side (after weak focusing), thus exciting guided modes in the patterned surface by virtue of the circular coupling slit. The aperture-less tip of the microscope scatters the electric near-field component that is normal to the surface onto a parabolic mirror, and its relative phase is extracted via interference with a temporally-modulated reference beam. (**d**),(**e**) Amplitude and phase measurement of a patterned surface with p=250nm and h=34nm, respectively. (**f**) Fourier-transform intensity of the complex field described by (d) and (e) inside the red dotted box, representing the reciprocal space. (**g**) Simulation of the spatial dispersion in the same patterned surface as in (d)-(f), including losses in the metal (supplement section 1). The colored rectangles represent different BZs: red, green and orange correspond to the first, the second and the third BZs, respectively.

Fig. 4 depicts experimentally measured EFCs of various slack metasurfaces with different geometrical parameters, so as to demonstrate the control over the surface waves guided along the metasurface. Fig. 4a1-3 provides representative EFCs obtained for small periodicities (140nm) at different modulation depths. Evidently, the circular EFC of the unperturbed surface changes to elliptic at shallow modulation and flips its curvature to a hyperbolic EFC at a deeper modulation, as predicted by the EMT and simulated in fig. 2. Fig. 4b1-5 depicts depth-dependent transitions, demonstrated for a constant periodicity of 210nm (~$\lambda$/3). Notably, the proximity of the neighboring BZ modifies the transition, boosting the gap opening at shallow modulation (47nm) to form the parabolic dispersion [27]. Deeper modulation gradually flattens the curve into a flat EFC (100nm depth) and then flips the curvature into hyperbolic (116nm depth). Finally, fig. 4c1-4 presents the periodicity-dependent topological double-transition, wherein, at a roughly constant depth of 50nm, the gap closes and reopens for increasing periodicity. The geometrical parameters associated with the EFCs present are marked in the "phase-diagram" that is re-plotted in fig. 4d along with the transitions between different propagation regimes. The complete measurement set is shown in the supplement section 4, along with the corresponding analytical calculations, exhibiting excellent agreement with the experimental results.

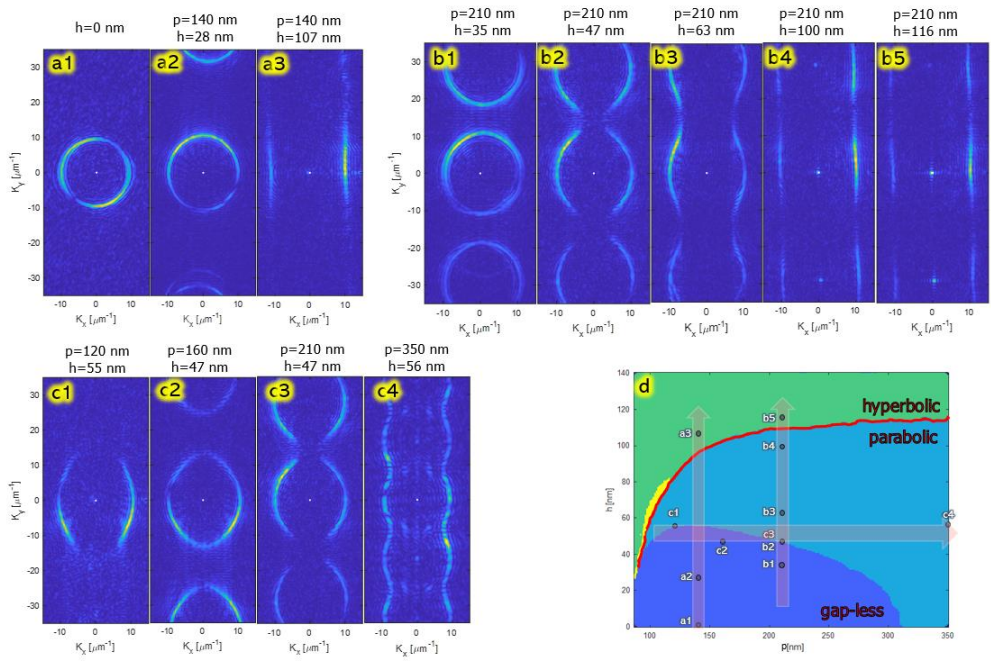

Figure 4- **Experimental demonstration of dispersion control and topological transitions in slack metasurface**. **(a-c)** equi-frequency contours for different structural parameters, extracted from the near-field measurement, labeled with their periodicity (letter p) and depth (letter h). All plots correspond to groove width of 80 $nm$ and incident wavelength of 660 $nm$. The origin of the reciprocal space is marked with a white dot in all plots. **(a1-3)** depth-dependent transition at dense metasurfaces (140nm periodicity). The EFC changes from being circular (a1) at a smooth surface, to elliptic (a2) at shallow modulation and hyperbolic (a3) at deep modulation. This observation is consistent with the EMT and with previous report [15]. **(b1-b5)** depth-dependent transition beyond effective medium; The EFC transforms from a series of closed (gapless) curves (b1) to an open parabolic "phase", mediated by the proximity of the curves (b2,3). Further deepening the groves results in a flat (b4) and a hyperbolic (b5) EFC. **(c1-4)** periodicity-dependent double topological transition; for an increasing period, open parabolic curves (c1) become gapless (c2) due to smaller anisotropy and reopens at larger period due to BZ "merging" (c3). For a large period $p > \frac{\lambda}{2}$ the curves comply with the tight-binding approximation [28]. **(d)** The geometrical parameters of the metasurfaces shown in (a-c) overlayed on the "phase diagram" as shown in fig. 2b; each sub-figure is labeled and marked as a circle on the diagram. The transitions are marked as arrows corresponding to the groups of sub-figures: a, b or c.

## 4. Umklapp scattering of surface waves

The close proximity between curves of adjacent BZs, as apparent in fig. 3f and fig. 4b1, provides a rare opportunity to study coupling effects and momentum transfer among the different BZs, known as *umklapp scattering*. This phenomenon is unique to periodic media, where scattering induced by the periodicity relaxes the momentum of a particle into its lowest permitted momentum. This process has a great impact on materials properties, as it determines the maximal thermal [29] and electrical [30] conductivity of materials and influencing the diffraction of light and electrons in diffraction gratings [31–33] and in photonic crystals [34,35] as demonstrated in various wave systems [36,37]. Although this process may limit the performance of devices, as occurring in a phased antenna array [38], many applications profit such scattering as it also allows topological edge states [39], supports unique transitions for photoemission [40] and provide special nonlinear phase matching conditions [41].

We utilize the analysis shown above to design a structure with the appropriate set of parameters to benefit this process and demonstrate for the first time *umklapp scattering of surface waves.* The design includes a relatively large modulation period, so as to provide a minimal momentum relaxation from the second to the first BZ, yet sufficiently small to preserve the gap-less EFC required to demonstrate umklapp scattering. Therefore, we select a configuration that features closed curves that are relatively close to each other. We further construct a coupling mechanism to excite surface waves through high BZ wavevectors only (see supplement section 5) and track their propagation characteristics using near-field microscopy. While recent theoretical studies [34–36] only considered umklapp scattering of a single wavevector (corresponding to a plane-wave), here we demonstrate an umklapp process of a two-dimensional beam consisting of a continuous set of in-plane wavevectors.

Fig. 5 demonstrates this unique property, exhibiting backward focusing as a result of umklapp surface-scattering. Figure 5a displays the equi-frequency contour, shown in fig. 3f, where the spatial constituents we aim to excite reside in the half-circles marked by the orange rectangles. We carve a specially-designed lens-shaped grating coupler over the modulated surface, shown in fig. 5b, to optimally couple a weakly focused laser beam into a continuous wavevectors set that resides in the second BZ. Note that this coupler excites both forward-propagating (upward in the figure) and backward-propagating (downward) constituents (supplement section 5). Figure 5c,d depict the near-field scanning microscope measurement of the normal component of the electric field phasor ($E_z$) across the area shown in fig. 5b. This engineered coupling results in a focal spot in the outer side of the lens-shaped coupler, corresponding to anomalous focusing that has so far been observed only in hyperbolic media [42] and in photonic crystals with concave bands [43]. For comparison, applying similar coupling grating to a smooth metal-dielectric interface (flat, unmodulated surface) shows no focusing on that side of the grating (supplement section 6).

In extracting the Fourier transform of the normal electric field component (fig. 5e), we find that the coupled Bloch mode contains predominately wavevectors in the *first* BZ (upper half) rather than in the second BZ which is targeted by the coupler. It indicates that the anomalous focusing observed is a result of umklapp scattering, in contrast to hyperbolic focusing of polaritons [42] or a photonic crystal negative refraction [44,45]. Our analysis facilitates digital Fourier filtering by which we visualize the projection of the Bloch modes on the plane wave basis, resulting in the contribution of different BZs (fig. 5f). We find that the amplitude profiles are similar for all BZs, while the phase-fronts are opposite both in their curvature and their evolution. Specifically, fig. 5f1 clearly shows normal focusing, associated with the first BZ and forward phase propagation, while fig. 5f2 depicts hyperbolic-like (concave) phase front propagating in the direction opposite to the energy flow which stems from the contribution of the second BZ (which possess here only negative wavevectors) to the Bloch mode.

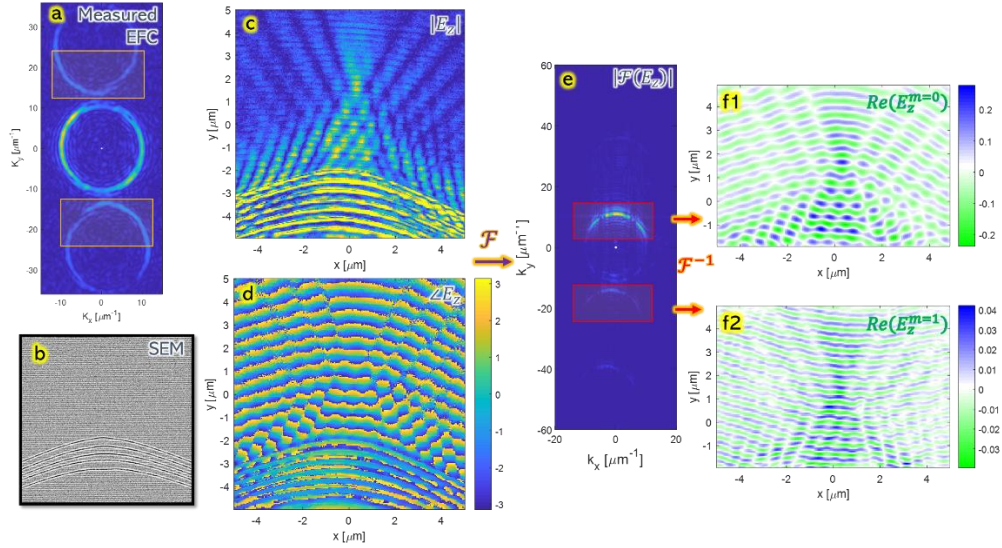

Figure 5- **Umklapp-mediated anomalous focusing**. **(a)** The equi-frequency contour mapping the modes of a surface modulated at periodicity 250nm at 34nm depth, acquired within a circular coupling slit (similar to fig. 3f). **(b)** SEM micrograph of the modulated surface and the concave-shaped coupling grating, designed to selectively couple free-space light into the $2^{nd}$ BZ, marked by the orange rectangles in (a). **(c),(d)** Near field measurement of the normal electric field phasor across this area: (c) amplitude, (d) phase. **(e)** A spatial frequency space image, extracted from the $E_z$ map shown in (c),(d). Note that while the lower red rectangle indicates the excited modes marked in fig 1a, the constituents contained in the upper red rectangle are not directly excited by the grating but a result of the umklapp scattering. **(f)** An inverse Fourier transform of the: forward-propagating zero order (f1) and the backward-propagating first order (f2) of the Bloch modes filtered separately; The colormap represents the real part of the phasor. (f1) shows normal propagation, characterized by convex phase fronts and a phase accumulated upwards. (f2) exhibit anomalous propagation, identified by the concave phase fronts, while the phase is accumulated downward.

Note that the anomalous diffraction shown here is not a collective property of the entire field, but rather represents a particular Fourier constituent. As such, it can be implemented in other periodic systems, without the need for special dispersion engineering as in photonic crystals and HMS or exotic diffraction as in Van-der Waals materials.

## 5. Conclusions

To conclude, our work introduced the concept of slack metsurfaces – artificial periodic media showing new schemes for surface waves propagation, beyond the effective medium approximation. Using a slack metasurface produced on a gold-air interface, we demonstrated the first instance of umklapp scattering in surface waves, exhibiting backward focusing of Bloch modes. We further showed how fully utilizing the structural degrees-of-freedom of modulated interfaces tunes their surface mode dispersion, resulting in unique topological transitions that natural or effective materials cannot manifest.

This concept can be readily implemented to shape wave propagation of other guided waves, including surface phonon-polaritons [46] and flying exciton-dipolaritons [47], each exhibiting a different "phase diagram" than the one shown in this study, potentially producing new and exotic wave dispersion regimes. Of particular importance is the capability we showed to engineer a double transition between a gapped and gapless EFC, as the interface between two gapped EFC of different topological origin should support the formation of robust edge states [48]. Moreover, even a continuous tunability of these transitions may be achieved in future works; while live control over periodicity is achievable either thermally or by using a piezoelectric substate voltage control, modulation depth can be effectively modified either by wavelength tuning or by a refractive index control (e.g., tuning voltage in liquid crystals), leading to a wide range of possible parameter regimes [49].

The umklapp focusing demonstrated herein, may pave the way for negative refraction in normal materials. While our demonstration currently accounts for negative refraction only for the second BZ, a special surface modulation with a modified unit cell, such as blazed gratings [50], may suppress all other harmonics of the guided mode such that the whole mode would become negatively refracting.

Extending the surface modulation demonstrated here to two-dimensions can further increase control over the spatial mode dispersion for applications the likes of all-angular diffraction-less propagation [45,51], while including several spatial frequencies [52] or modulating a multimode waveguiding system [53] may elicit intriguing inter-dependent umklapp processes, analogous to those occurring in moiré hyperbolic metasurfaces [18]. Such extensions to our work along with its potential compatibility with photolithography fabrication, may find uses in contemporary photonic circuits, paving the way to umklapp-inspired photonic components and devices.

## 6.   Methods

For sample fabrication, we deposit a 200nm gold layer upon a NBK7 glass substrate using e-Gun evaporation (Airco Temescal FC-1800, deposition rate: $5 \, Å/sec$) and pattern the slits of the slack metasurface and the couplers using a focused ion beam (FEI Helios NanoLab DualBeam G3 UC, accelerating voltage: 30KV, current: 7.7pA, home-made software). Afterwards, we thermally evaporate 10nm thick $SiO2$ layer upon the patterned gold surface (Vinci PVD-4, deposition rate: $0.5 \, Å/sec$) to prevent damage by the near-field probe. The signal from the near-field microscope (neaSNOM, Neaspec) is filtered spatially (using 4F filtering) and temporally (using pseudo-heterodyne technique [54], benefitting the third harmonic of tip oscillations).

**Funding.** This work was supported by the Israel Science Foundation (ISF) grant number 1750/18 and the Russel Berrie Nanotechnology Institute (RBNI) at the Technion.

**Disclosures.** The authors declare no conflicts of interest.

**Data availability.** The data that support the plots within this paper and other findings of this study are available from the corresponding author upon reasonable request.

**Supplemental document.** See Supplement Sections 1-6 and Supplemental Video 1-2 for supporting content.

## Manuscript References

# Topological transitions and surface umklapp scattering in Slack Metasurfaces: supplemental document

### Section 1: simulating the response in reciprocal space ('K-space')

The equi-frequency contour – the system's response in momentum space, is calculated using GD-calc [1]: an analytic Matlab simulation based on "Fourier Modal Method" (FMM) [2], calculating the diffraction orders from periodic structures with a transfer matrix. The simulation is provided with the structural dimensions and the zero-order reflection is calculated for a two-dimensional grid of lateral wave-vectors outside the light cone. High reflection coefficient points represent good surface coupling, therefore constructs the EFC.

Gold permittivity is taken from reference [3]; The presence of the glass substrate was neglected for slits shallower than $h < 130\ nm$, where the metallic properties screen any substrate effect.

### Section 2: Reconstruction of the equifrequency contours

The equifrequency contour shown in fig. 2 in the manuscript is simulated equivalently, while artificially decreasing the metallic losses to obtain a sharp resonance and calculating high diffraction orders (m=20) to eliminate a "flickering" artifact. The algorithm first detects the intersection of the EFC with the x-axis ($k_x = 0$), then recursively track the adjacent peak value until constructing the whole curve.

### Section 3: Preparation of the topological phase diagram

To construct the phase diagram, we calculate the EFC for each parameters set $(h, p)$ and define criteria for each phase, based on the contour points $\{k_x\}, \{k_y\}$, the intersection point $(k_{x_0}, 0)$ and the derivative $\left.\frac{dk_x}{dk_y}\right|_{(k_{x_0}, 0)}$.

The conditions for each phase (rows) appear in the table S1:

**Table S1- Conditions for each phase in the topological phase diagram**

|  | Condition 1 | Condition 2 | Condition 3 |
|---|---|---|---|
| **Gap-less** | $k_{x_0} = \max(\{k_x\})$ | $\left.\frac{dk_x}{dk_y}\right|_{(k_{x_0}, 0)} < 0$ | $\min(\{k_x\}) = 0$ |
| **Parabolic** | $k_{x_0} = \max(\{k_x\})$ | $\left.\frac{dk_x}{dk_y}\right|_{(k_{x_0}, 0)} < 0$ | $\min(\{k_x\}) > 0$ |
| **Hyperbolic** | $k_{x_0} = \min(\{k_x\})$ | $\left.\frac{dk_x}{dk_y}\right|_{(k_{x_0}, 0)} > 0$ | |
| **Hybrid** | $\min(\{k_x\}) < k_{x_0} < \max(\{k_x\})$ | $\left.\frac{dk_x}{dk_y}\right|_{(k_{x_0}, 0)} < 0$ | |

### Section 4: Comparing measurements to analytical simulations

Due to losses and merging between adjacent BZs, the spatial dispersion cannot be always described as a well-defined contour as singular points, gaps and wide bands of uncertain momentum break the line continuity. Consequently, the spatial modes are better illustrated as a two-dimensional surface-coupling map, as described in Section 1. In table 2, we compare our results from fig. 4 (left column) to the surface coupling map we extract from the FMM simulation (right column). Indeed, we obtain an excellent agreement with the theory; sharp curves in the simulation (correspond to small uncertainty in the momentum; for example, all points in a1,a2,b1,c2 and c4) are clearly distinguished in the experiments, while broad distribution in momentum space, that represent high uncertainty appear much weaker in the experiments (for example, the gap in b2, c1 and in c2, points close to the BZ boundary in a3, b3 b4 and in b5).

**Table S2- Comparison of measurements to losses counting simulation.**

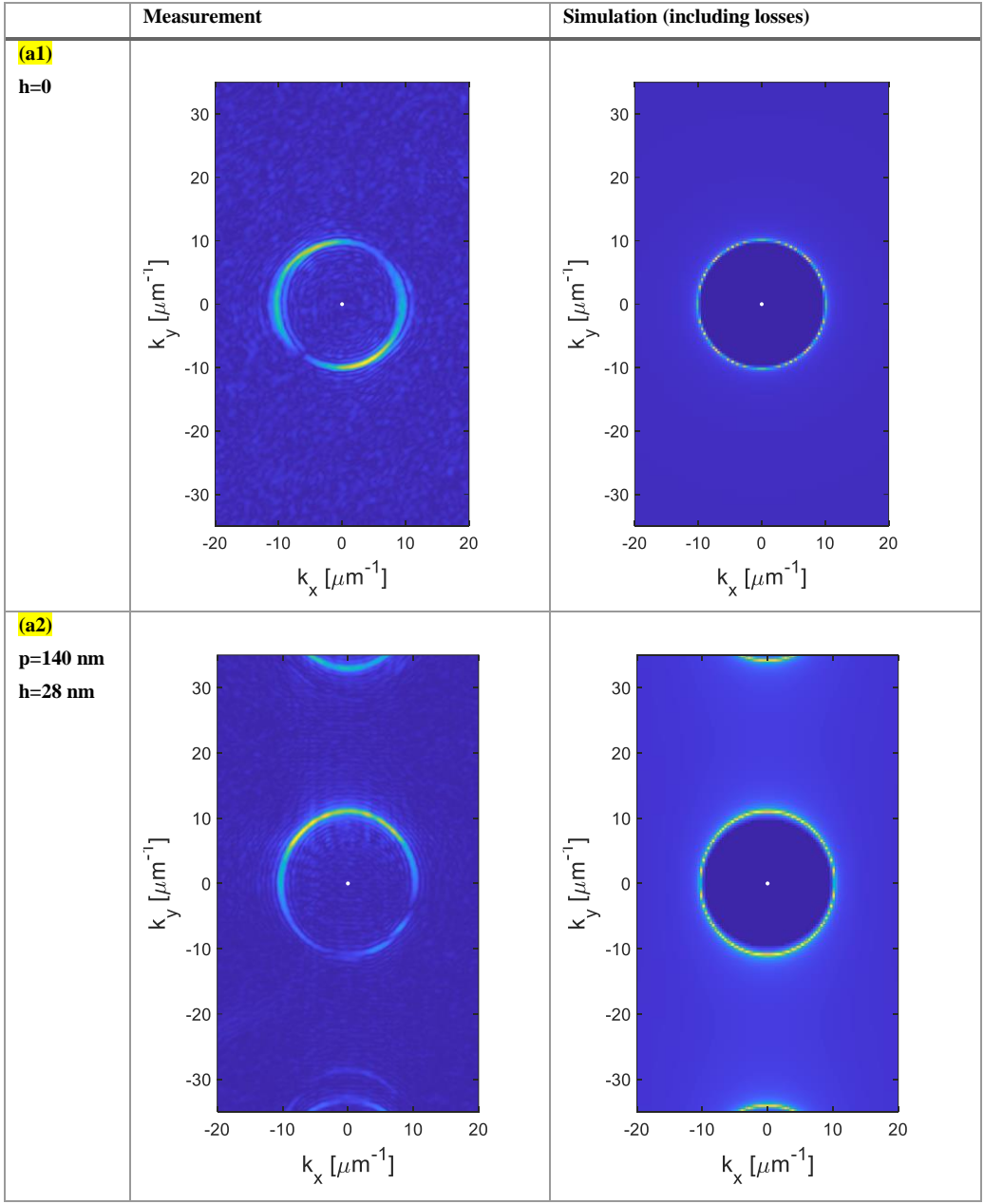

| | Measurement | Simulation (including losses) |
|---|---|---|
| **(a1)** h=0 | | |
| **(a2)** p=140 nm h=28 nm | | |

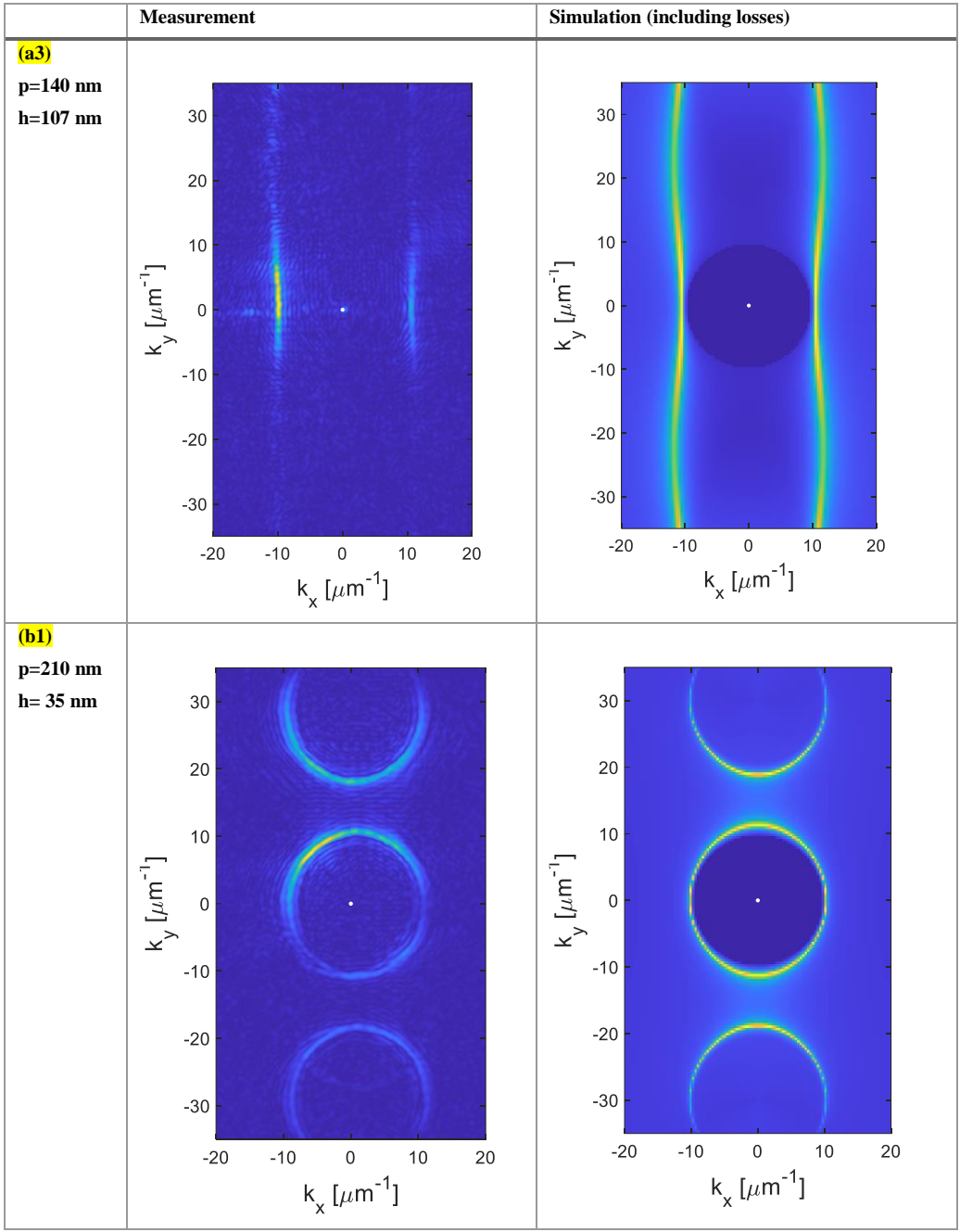

| | Measurement | Simulation (including losses) |
|---|---|---|
| **(b2)** **(c3)** p=210 nm h= 47 nm |  |  |
| **(b3)** p=210 nm h= 63 nm |  |  |

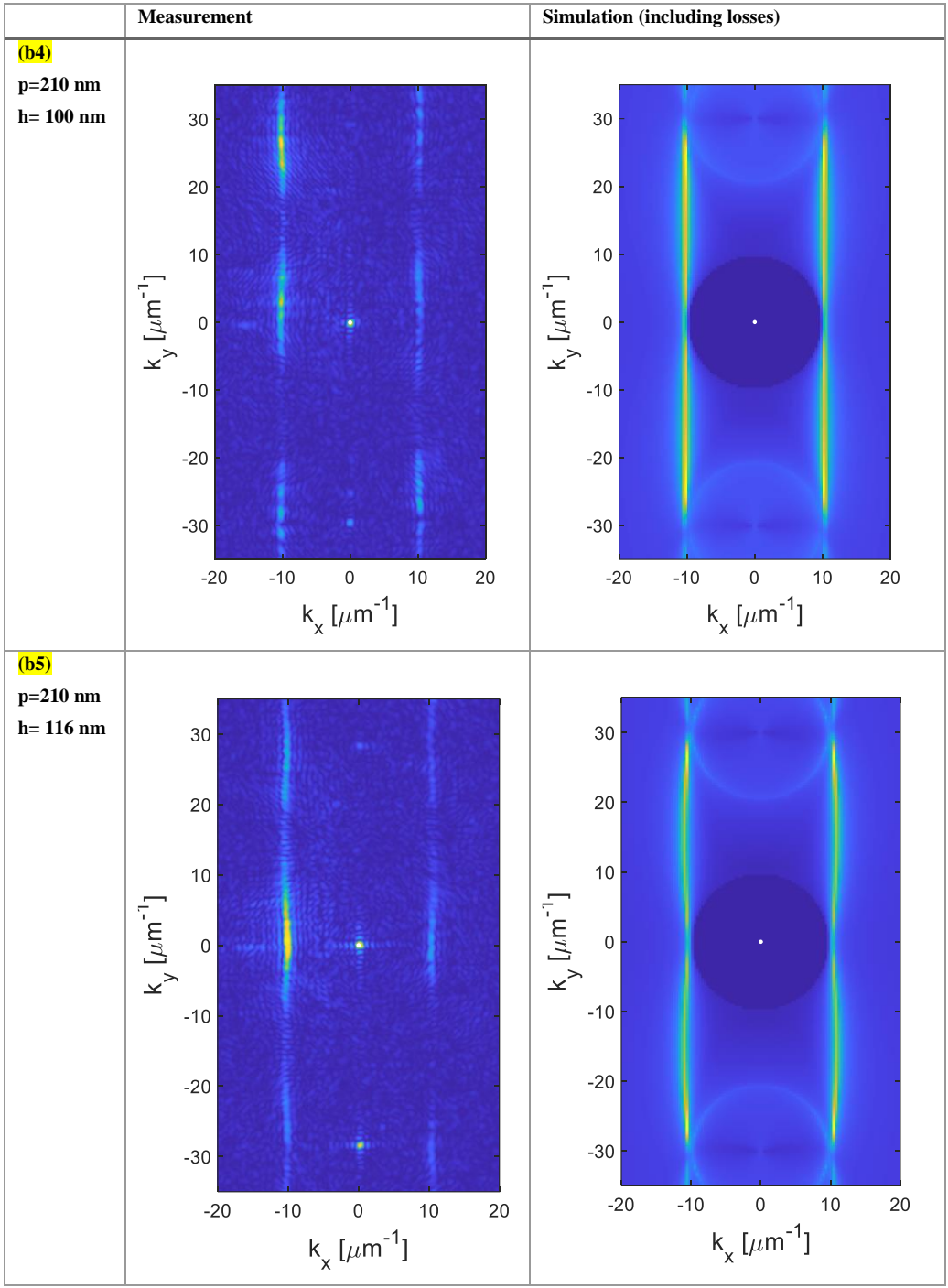

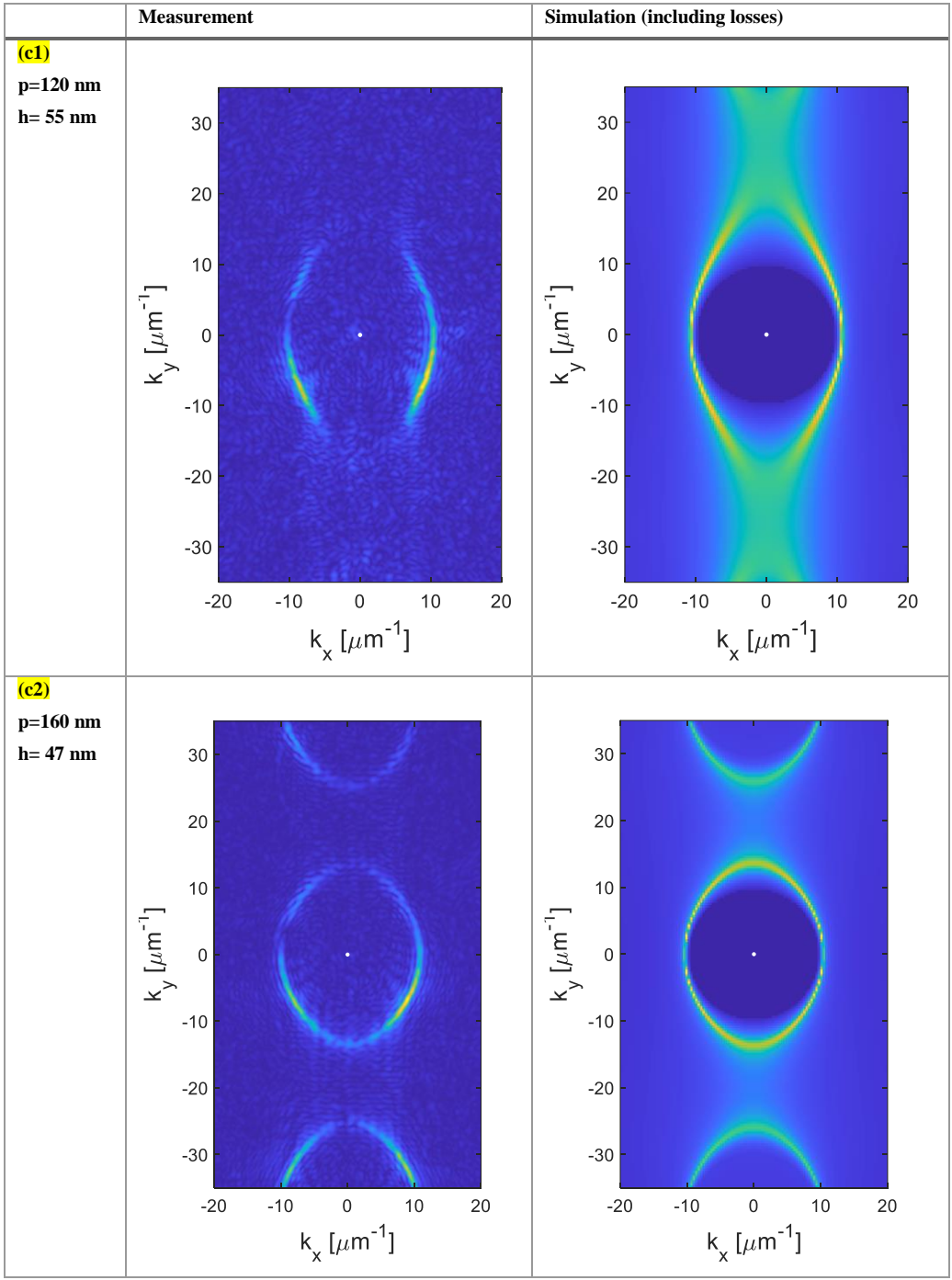

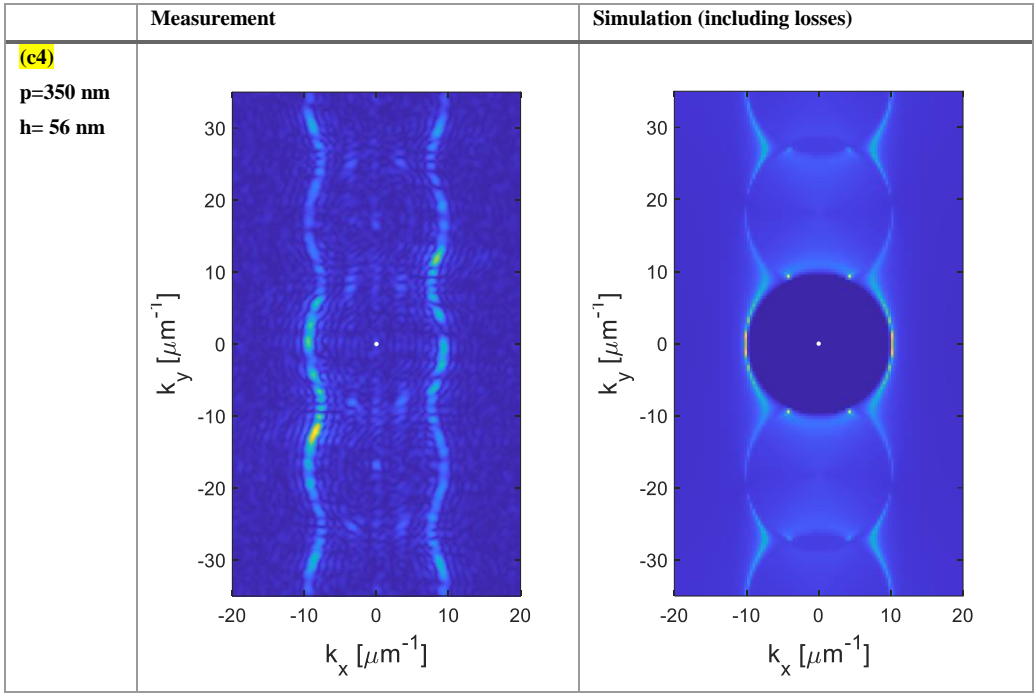

| | Measurement | Simulation (including losses) |
|---|---|---|
| **(c4)** **p=350 nm** **h= 56 nm** | | |

### Section 5: Design of the backward focusing coupler

We design a backward surface-waves focuser, mediated by excitation of the second BZ utilizing reciprocal problem: we situate a time-harmonic impulse at the focal spot and track its equi-phase lines as it propagates towards the desired coupler location. Since the momentum space of the impulse response is calculated using FMM, the real-space propagation of the impulse can be calculated using an inverse Fourier transform.

Figure S1 illustrates the design process of the couplers; we compute the momentum distribution using FMM (fig. S1a), filter only the lower half band of the second BZ (Fig.S1b) and obtain the impulse response using a Fourier transform (Fig. S1c,d). Since the surface is illuminated by a constant planar phase, the red-marked couplers generate a wavepacket that is "time-reversed" with respect to the calculated impulse response hence results in focusing back into the calculated focal spot. We apply an arbitrary global phase for the equi-phase contour and pattern multiple coupling slit along these lines to improve the coupling process and to avoid undesired coupling to other bands in a similar way to multimode waveguides. Since the length of the equi-phase contour (marked in red in fig. S1d) is infinite, we construct the coupling slits using only the segment bounded in $|x| < 10\mu m$.

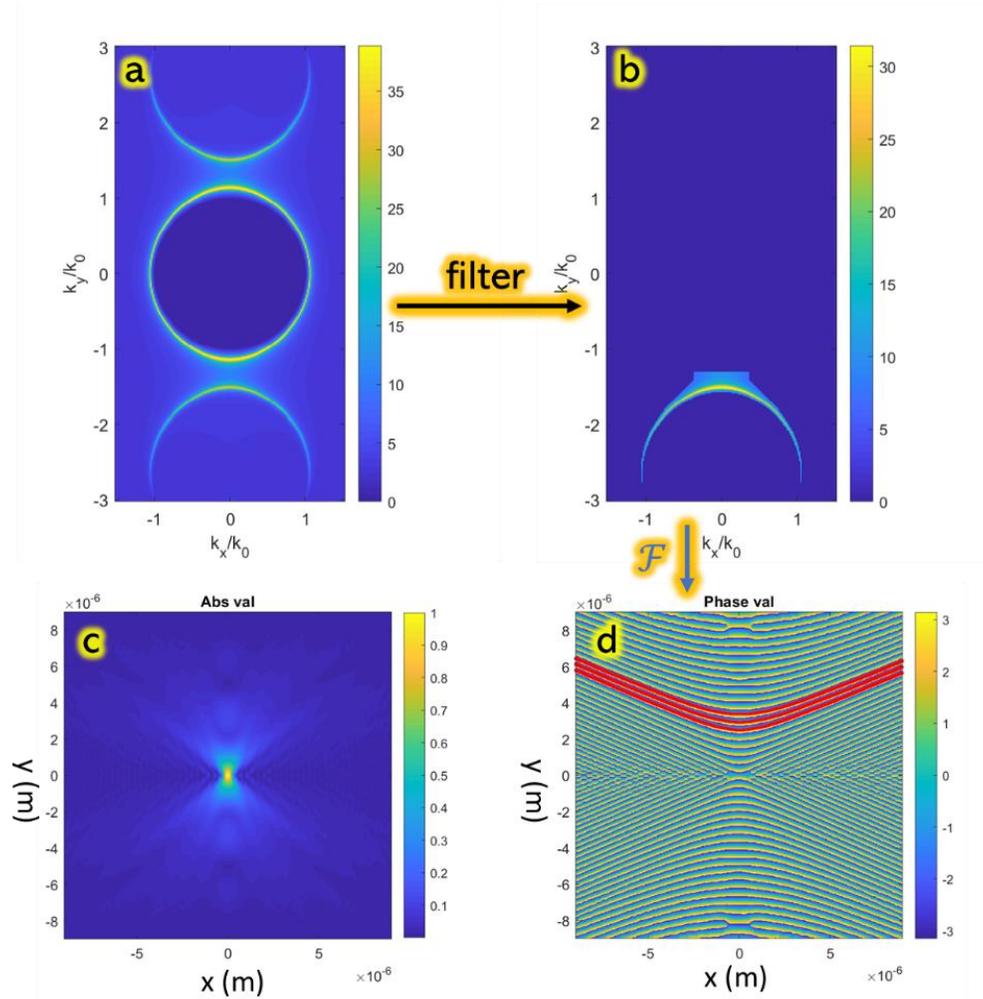

Figure S1- **process of coupler design**. **(a)** zero-order reflection coefficient map, attained with high resolution for the Fourier transform smoothness. **(b)** second BZ filtering. **(c)** normalized amplitude and **(d)** phase images of the Fourier transform of b. Three equi-phase contours are patterned as couplers, marked in red.

In this design, we select the band of negative $k_y$, yet the equivalent positive $k_y$ band could be selected as well; these bands are solely different in the phase accumulation direction and the energy flow direction, respectively. Practically, each side of the slit transmits waves with a different sign of $k_y$.

### Section 6: Backward focusing compared to a reference measurement

To verify that the anomalous backward focusing emanates directly from umklapp process, we performed a reference measurement of the same coupling slits without a slack metasurface, fig. S2. While fig. S2a,b (same as fig. 5c,d) demonstrate anomalous focusing and its Fourier transform respectively, fig. S2c,d show the matching reference measurement. Evidently, no focusing is achieved in the reference measurement and the numerical aperture of the excitation (namely, the opening angle of the excited modes in the momentum space) is significantly smaller.

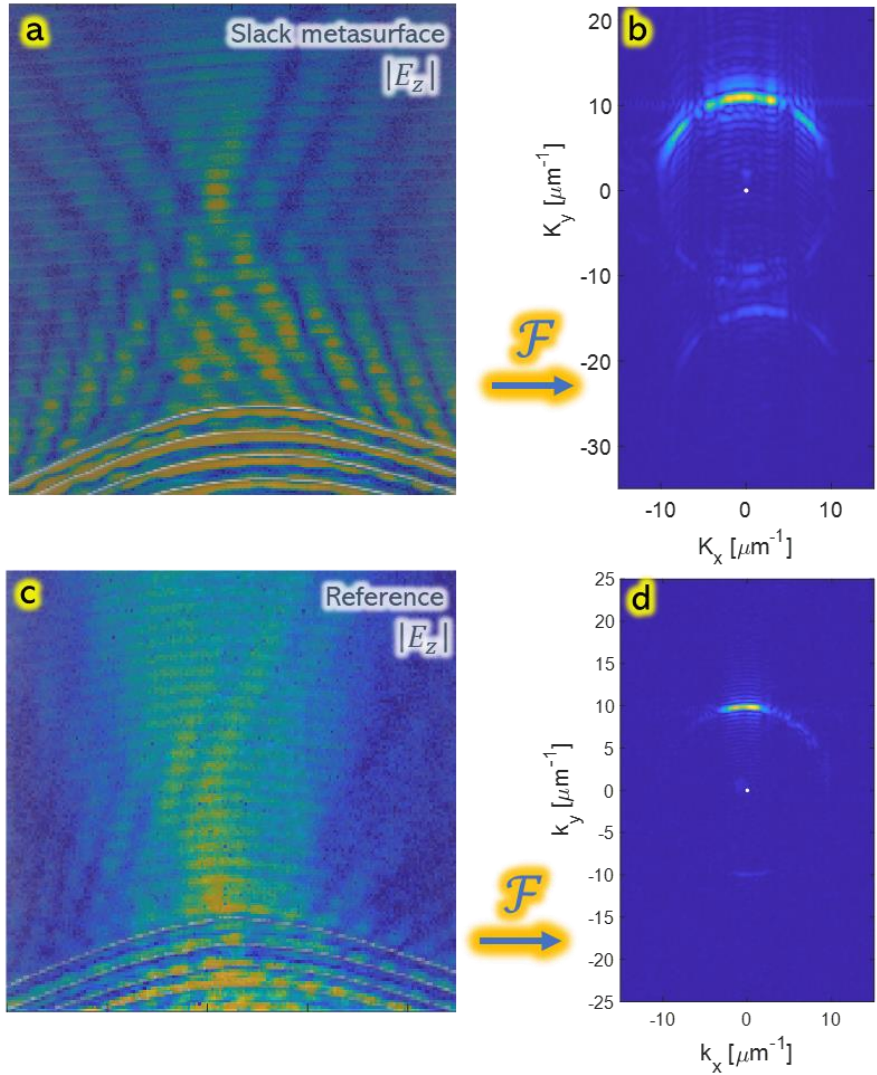

Figure S2- **Umklapp-driven backward focusing measurement compared to a reference measurement**. **(a)** Measurement of a backward focusing, identical to Fig. 5c, overlayed with the AFM image. The yellow-blue scale corresponds to the amplitude of the perpendicular E-field phasor (yellow is maximum), while the black-white scale corresponds to the depth profile (white is the deepest). Here, a slack metasurface of 250nm periodicity and 34nm depth is utilized for umklapp scattering originating in the second BZ. **(b)** The corresponding $E_z$ Fourier transform, same as fig. 5e. **(c)** Measurement of $E_z$ generated by the same coupling slits carved on a flat gold surface, overlayed with AFM image. Here, a backward focusing is not apparent. **(d)** The corresponding Fourier transform, showing a much narrower numerical aperture. This illustrates the capability of surface-umklapp process to scatter a broad range of wave-vectors.

## Supplementary References